\documentclass[12pt]{article}

\usepackage{a4}
\usepackage[pdftex,usenames,dvipsnames]{color}
\usepackage{graphics}
\usepackage{amsfonts}

\newcommand{\nc}{\newcommand}
\nc{\blue}{\textcolor{Blue}}
\nc{\navy}{\textcolor{NavyBlue}}
\nc{\cyan}{\textcolor{Cyan}}
\nc{\violet}{\textcolor{Violet}}
\nc{\purple}{\textcolor{Purple}}
\nc{\red}{\textcolor{Red}}
\nc{\sienna}{\textcolor{RawSienna}}
\nc{\brown}{\textcolor{Brown}}
\nc{\orange}{\textcolor{Orange}}
\nc{\yellow}{\textcolor{Yellow}}
\nc{\green}{\textcolor{Green}}
\nc{\olive}{\textcolor{OliveGreen}}

\newcommand{\nit}{\noindent}

\newcommand{\np}{\newpage}
\newcommand{\dsp}{\displaystyle}
\newcommand{\vs}[1]{\vspace{#1 ex}}
\newcommand{\hs}[1]{\hspace{#1 em}}
\newcommand{\bflr}{\begin{flushright}}
\newcommand{\eflr}{\end{flushright}}
\newcommand{\bc}{\begin{center}}
\newcommand{\ec}{\end{center}}
\newcommand{\ben}{\begin{enumerate}}
\newcommand{\een}{\end{enumerate}}

\newcommand{\be}{\begin{equation}}
\newcommand{\ee}{\end{equation}}
\newcommand{\ba}{\begin{array}}
\newcommand{\ea}{\end{array}}
\newcommand{\ld}{\left.}
\newcommand{\rd}{\right.}
\newcommand{\ct}{\cite}
\newcommand{\bit}{\bibitem}
\newcommand{\dd}[2]{\frac{\partial{#1}}{\partial{#2}}}

\newcommand{\del}{\delta}

\newcommand{\ve}{\varepsilon}

\newcommand{\thg}{\theta}
\newcommand{\kg}{\kappa}
\newcommand{\lb}{\lambda}

\newcommand{\rg}{\rho}

\newcommand{\vf}{\varphi}
\newcommand{\og}{\omega}
\newcommand{\Gam}{\Gamma}
\newcommand{\Del}{\Delta} 

\newcommand{\Fg}{\Phi}

\newcommand{\Lb}{\Lambda}
\newcommand{\lh}{\left(}
\newcommand{\rh}{\right)}

\newcommand{\nb}{\nabla}

\newcommand{\cD}{{\cal D}}

\begin{document}
\pagestyle{empty}

\bflr
NIKHEF/2014-27
\eflr

\bc
\Large{ \bf Gauge-covariant extensions of Killing tensors}
\vs{2}

\Large{ \bf and conservation laws}
\vs{5}

\large{J.W. van Holten$^a$} 
\vs{2}

\large{Nikhef }
\vs{1}

\large{ Science Park 105, Amsterdam NL}
\vs{1}

\large {and}
\vs{1}

\large{Lorentz Institute, Leiden University}
\vs{1} 

\large{Niels Bohrweg 2, Leiden NL}
\ec
\vs{3}

\nit
{\small {\bf Abstract} \\
In classical and quantum mechanical systems on manifolds with gauge-field fluxes, constants of motion are constructed from 
gauge-covariant extensions of Killing vectors and tensors. This construction can be carried out using a manifestly covariant 
procedure, in terms of covariant phase space with a covariant generalization of the Poisson brackets, c.q. quantum 
commutators. Some examples of this construction are presented.}

\vfill
\footnoterule
\nit
{\footnotesize $^a$ e-mail: v.holten@nikhef.nl}

\np
\pagestyle{plain}
\pagenumbering{arabic}

\section{Noether's theorem}

This paper discusses symmetries and conservation laws in the context of hamiltonian dynamics. The discussion is framed
predominantly in the language of classical dynamics, but the use of Poisson brackets and their correspondence with 
quantum commutators,  guarantees that many results also apply to the operator formulation of quantum dynamics. The 
main difference is the operator ordering to be implemented in quantum theory, the technicalities of which are not relevant 
to the issues I focus on. 

The connection between continuous symmetries and conservation laws is established by Noether's theorem 
\ct{noether1918}. I briefly review the theorem by considering infinitesimal transformations on phase-space variables 
$(x, p)$ obtained from a generating function $G(x,p)$ through the Poisson brackets 
\be
\del x = \left\{ x, G \right\} = \dd{G}{p}, \hs{2} \del G = \left\{ p, G \right\} = - \dd{G}{x}.
\label{1.1}
\ee
Observe, that these variations are defined such that $G$ itself is an invariant:
\be
\del G = \del x \dd{G}{x} + \del p \dd{G}{p} = \left\{G, G \right\} = 0.
\label{1.2}
\ee
Under such transformations the hamiltonian of the system changes by
\be
\del H = \left\{ H, G \right\} = - \frac{dG}{dt},
\label{1.3}
\ee
the change of $G$ along the phase-space trajectory $(x(t), p(t))$ generated by the hamiltonian $H$.
It follows immediately, that $G$ is a constant of motion if the hamiltonian is invariant under the
transformations (\ref{1.1}).

\nit
It is also of some interest to consider the variation of the action 
\be
S = \int_1^2 dt \lh p\, \frac{dx}{dt} - H(x,p) \rh.
\label{1.4}
\ee
Applying the variations (\ref{1.1}) 
\be
\del S = \int_1^2 dt \left[ \frac{d}{dt} \lh p\,\dd{G}{p} - G \rh - \left\{ H, G \right\} \right] = \left[ p \dd{G}{p} - G \right]_1^2.
\label{1.5}
\ee
Thus variations under which the hamiltonian is invariant, leave the action invariant modulo boundary terms. 
This is sufficient for $G$ to be a constant of motion.

\section{Isometries of manifolds}

On a manifold with (local) co-ordinates $x^{\mu}$ and metric $g_{\mu\nu}(x)$ the geodesics can be obtained 
as the trajectories of test-particles with proper-time hamiltonian
\be
H = \frac{1}{2}\, g^{\mu\nu} p_{\mu} p_{\nu}.
\label{2.1}
\ee
Indeed, using the overdot notation for proper-time derivatives, the hamilton equations take the form
\be 
\dot{x}^{\mu} = g^{\mu\nu} p_{\nu}, \hs{2} \dot{p}_{\mu} = - \frac{1}{2}\, \dd{g^{\nu\lb}}{x^{\mu}}\, p_{\nu} p_{\lb} 
 = \frac{1}{2}\, \dd{g_{\nu\lb}}{x^{\mu}}\, \dot{x}^{\nu} \dot{x}^{\lb}.
\label{2.2}
\ee
The last expression is equivalent to the geodesic equation
\be
\ddot{x}^{\mu} + \Gam_{\lb\nu}^{\;\;\;\mu}\, \dot{x}^{\lb} \dot{x}^{\nu} = 0.
\label{2.3}
\ee
In this language, isometries of the manifold are found as constants of motion which are linear in the momentum:
\be
J(x,p) = J^{\mu}(x) p_{\mu}, \hs{2} \left\{ J, H \right\} = \lh \nb_{\mu} J_{\nu} \rh p^{\mu} p^{\nu} = 0,
\label{2.4}
\ee
where (in a somewhat hybrid notation) the contravariant components of the momentum are $p^{\mu} = \dot{x}^{\mu}$.
Hence the covariant coefficient functions $J^{\mu}$ form a Killing vector, a  solution of the Killing equation
\be
\nb_{\mu} J_{\nu} + \nb_{\nu} J_{\mu} = 0 \hs{1} \Leftrightarrow \hs{1} 
J^{\lb} \dd{g_{\mu\nu}}{x^{\lb}} + \dd{J^{\lb}}{x^{\mu}}\, g_{\lb\nu} + \dd{J^{\lb}}{x^{\nu}}\, g_{\mu\lb} = 0. 
\label{2.5}
\ee
The second (contravariant) form of the equation states that the Lie-derivative of the metric w.r.t.\ the vector $J^{\mu}$
vanishes, which is the usual definition of an isometry. Also note, that the constants of motion defined by Killing
vectors are precisely those, for which
\be
p\, \dd{G}{p} = G,
\label{2.6}
\ee
and therefore generates transformations under which  the action is strictly invariant. This is to be expected for an
isometry which by construction leaves the line element $ds^2 = g_{\mu\nu} dx^{\mu} dx^{\nu}$ invariant. 

Although the coordinate transformations $\del x^{\mu}$ generated by Killing vectors thus have an elegant interpretation,
this does not hold for the corresponding transformations of the canonical momenta $p_{\mu}$:
\be
\del p_{\mu} = A_{\mu}^{\;\nu}(x) p_{\nu}, \hs{2} A_{\mu}^{\;\nu} = - \dd{J^{\nu}}{x^{\mu}}.
\label{2.7}
\ee
This transformation rule is not general covariant; as $A_{\mu}^{\;\nu}(x)$ is point-dependent, covariance requires 
$\del p$ to be corrected for the parallel displacement generated by the translation $\del x^{\mu}$. Thus a set of 
covariant transformations in phase space is defined by
\be
\Del x^{\mu} = \del x^{\mu}, \hs{2} \Del p_{\mu} = \del p_{\mu} - \del x^{\lb} \Gam_{\lb\mu}^{\;\;\;\nu} p_{\nu}.
\label{2.8}
\ee
In order for these transformations to respect the Poisson brackets (\ref{1.2}) and (\ref{1.3}), it is then necessary
to introduce a covariant derivative
\be
\cD_{\mu} G = \dd{G}{x^{\mu}} + \Gam_{\mu\nu}^{\;\;\;\lb}\, p_{\lb}\, \dd{G}{p_{\nu}} = - \Del p_{\mu},
\label{2.9}
\ee
such that
\be
\Del G = \Del x^{\mu} \cD_{\mu} G + \Del p_{\mu} \dd{G}{p_{\mu}} = \left\{ G, G \right\} = 0,
\label{2.10}
\ee
and
\be
\Del H = \Del x^{\mu} \cD_{\mu} H + \Del p_{\mu} \dd{H}{p_{\mu}} = \left\{ H, G \right\} = - \frac{dG}{d\tau}.
\label{2.11}
\ee
Thus we have constructed a covariant expression for the Poisson bracket of two arbitrary scalar phase-space 
functions:
\be
\left\{ G, K \right\} = \cD_{\mu} G\, \dd{K}{p_{\mu}} - \dd{G}{p_{\mu}}\, \cD_{\mu} K.
\label{2.12}
\ee

\section{Killing tensors}

The metric postulate guarantees that the geodesic hamiltonian (\ref{2.1}) is covariantly constant:
\be
\nb_{\lb} g_{\mu\nu} = 0 \hs{1} \Leftrightarrow \hs{1} \cD H = 0. 
\label{3.1}
\ee
The condition for a constant of geodesic motion then takes the simple form
\be
\left\{ G, H \right\} = p^{\mu} \cD_{\mu} G = 0.
\label{3.2}
\ee
Applying this to a general expression of homogeneous rank $n$ in the momenta 
\be
G(x,p) = G^{\mu_1 ... \mu_n}(x) p_{\mu_1} ... p_{\mu_n} \hs{1} \Rightarrow \hs{1}
p^{\mu} \cD_{\mu} G = \lh \nb_{\mu_{n+1}} G_{\mu_1 ... \mu_n} \rh p^{\mu_1} ... p^{\mu_{n+1}},
\label{3.3}
\ee
the condition for a constant of motion becomes a generalization of the Killing equation (\ref{2.5}):
\be
\nb_{\lh \mu_{n+1} \rd} G_{\ld \mu_1 ... \mu_n \rh} = 0.
\label{3.4}
\ee
The solutions of these equations are therefore known as {\em Killing tensors} \ct{carter1977,vholten-rietdijk1993}. 

The geometrical interpretation of the transformations generated by constants of motion constructed from 
Killing tensors of rank 2 or higher, is more complicated than for Killing vectors. They do not generate 
transformations in the manifold, but in the tangent bundle, the physical phase-space:
\be
\ba{l}
\dsp{ \Del x^{\mu} = \dd{G}{p_{\mu}} = n G^{\mu\mu_1...\mu_{n-1}} p_{\mu_1} ... p_{\mu_{n-1}}, }\\
 \\
\dsp{ \Del p_{\mu} = - \cD_{\mu} G = - \lh \nb_{\mu} G^{\mu_1...\mu_n} \rh p_{\mu_1} ... p_{\mu_n}. }
\ea
\label{3.5}
\ee
Note, that for transformations with $n \geq 2$ the action actually transforms by boundary terms:
\be
p_{\mu}\, \dd{G}{p_{\mu}} = n G \hs{1} \Rightarrow \hs{1} 
\Del S = (n - 1) \left[ G(x_2,p_2) - G(x_1, p_1) \right].
\label{3.6}
\ee
Obviously, as $G$ is a constant of motion the action is still invariant along geodesics.
\vs{1}

\nit
{\em Example: Kerr geometry} \ct{carter1968,carter1977} \\
The Kerr geometry is a Ricci-flat geometry in 4-dimensional space-time: $R_{\mu\nu} = 0$, implying it is
a solution of the Einstein equations of General Relativity in empty space-time. This solutions describes 
a rotating black hole with constant mass $M$ and angular momentum $J$. The geodesic hamiltonian reads
\be
H  = \frac{1}{2\rg^2} \left[ \Del^2 p_r^2 + p_{\thg}^2 + \lh a \sin \thg\, p_t + \frac{p_{\vf}}{\sin \thg} \rh^2
 - \frac{1}{\Del^2} \lh (r^2 + a^2) p_t + a p_{\vf} \rh^2 \right], 
\label{3.7}
\ee
where $a = J/M$ is the angular momentum per unit of mass, and the notation follows standard conventions:
\[
\Del^2 = r^2 - 2Mr + a^2, \hs{2} \rg^2 = r^2 + a^2 \cos^2 \thg.
\]
This space-time geometry admits a rank-2 Killing tensor, coding the well-known Carter constant of motion:
\be
\ba{lll}
K & = & \dsp{ \frac{1}{2\rg^2} \left[ - \Del^2 a^2 \cos^2 \thg\, p_r^2 + r^2 p_{\thg}^2 + r^2 \lh a p_t 
 + \frac{p_{\vf}}{\sin^2 \thg} \rh^2 \rd }\\
 & & \\
 & & \dsp{ \hs{2.5} \ld +\, \frac{a^2 \cos^2 \thg}{\Del^2} \lh (r^2 + a^2) p_t + a p_{\vf} \rh^2 \right]. }
\ea
\label{3.8}
\ee
This constant of motion  can be generalized to neutral and charged, spinless and spinning particles in Kerr-Newman 
space-time of charged black holes \ct{carter1977,penrose1973, floyd1973,gibbons-rietdijk-vholten1993}. 

\section{Bracket algebra}

The generators of hamiltonian symmetry transformations, defining constants of motion, define a Lie-algebra by their 
bracket relations (\ref{2.12}); this follows from the Jacobi identity
\be
\left\{ \left\{ G, K \right\}, J \right\} + \left\{ \left\{ K, J \right\} , G \right\} + \left\{ \left\{ J, G \right\} , K \right\} = 0.
\label{4.1}
\ee
Applying the identity to the special case where one of the functions is the hamiltonian: $J = H$, implies that 
constants of motion $(G, K)$ satisfy
\be
\left\{ \left\{ G, K \right\}, H \right\} = 0.
\label{4.2}
\ee
Hence the bracket of two constants of motion produces another constant of motion, and 
the set of such constants is closed under the bracket operation. Here we summarize some properties
of this algebra \ct{vholten-rietdijk1993}. 

The generators $J = J^{\mu} p_{\mu}$ linear in momentum define a Lie subalgebra of the full algebra:
\be
\left\{ J_1, J_2 \right\} = J_3, \hs{2} J_3^{\mu} = J_2^{\nu}\, \nb_{\nu} J_1^{\mu} - J_1^{\nu}\, \nb_{\nu} J_2^{\mu}.
\label{4.3}
\ee
The constants of motion of rank $n \geq 2$ then define representations of this subalgebra\footnote{The parenthesis 
around the indices $(\mu_1 ... \mu_n)$ denote complete symmetrization with unit weight.}, characterized by the
transformation rule
\be
\left\{ G_1, J \right\} = G_2, \hs{2} G_2^{\mu_1 ... \mu_n} = 
  J^{\lb} \nb_{\lb} G^{\mu_1 ... \mu_n} - n G^{\lb \lh \mu_1 ... \mu_{n-1} \rd} \nb_{\lb} J^{\ld \mu_n \rh}.
\label{4.4}
\ee
The Jacobi identity 
\be
\left\{ \left\{ G, J_1 \right\}, J_2 \right\} - \left\{ \left\{G, J_2 \right\}, J_1 \right\} = \left\{ G, \left\{ J_1, J_2 \right\} \right\},
\label{4.5}
\ee
then guarantees that the transformations generated by the linear $J(x,p)$ on the rank-$n$ tensors $G(x,p)$ obey the composition
rules of the subalgebra spanned by the linear $J(x,p)$ themselves.

As for the brackets of a generator $G^{(n)}$ of rank $n$, and a generator $G^{(m)}$ of rank $m$, with both $(n,m) \geq 2$,
it is easily recognized that any non-vanishng brackets of such quantities must be a generator $G^{(n+m-1)}$, of
rank $n+m-1$ exceeding both $n$ and $m$:
\be
\left\{ G^{(n)}, G^{(m)} \right\} \sim G^{(n+m-1)}. 
\label{4.6}
\ee
Therefore non-vanishing brackets of higher-rank generators will potentially lead to an infinite set of generators of 
arbitrary high rank. Well-known examples of such algebras are the Virasoro and Kac-Moody algebras of 2-D conformal 
field theories.

\section{Abelian gauge interactions}

Geodesic motion on a manifold applies to the motion of a pure mass point subject only to geometrical forces. 
However, Noether's theorem is very general and can be applied also in presence of external force fields 
\ct{jackiw-manton1980,duval-horvathy1982}.  
In this section I consider abelian gauge interactions transmitted by a vector field $A_{\mu}(x)$, acting on a 
point mass $m$ with charge $q$. For such a particle the canonical hamiltonian reads
\be
H = \frac{1}{2m}\, g^{\mu\nu} \lh p_{\mu} - q A_{\mu} \rh \lh p_{\nu} - q A_{\nu} \rh.
\label{5.1}
\ee
The hamiltonian equations of motion (\ref{2.2}), (\ref{2.3}) are generalized to
\be
p_{\mu} = m g_{\mu\nu} \dot{x}^{\nu} + q A_{\mu}, \hs{2} 
g_{\mu\nu} \lh \ddot{x}^{\nu} + \Gam_{\kg\lb}^{\;\;\;\nu} \dot{x}^{\kg} \dot{x}^{\lb} \rh = \frac{q}{m}\, F_{\mu\nu}\, \dot{x}^{\nu},
\label{5.2}
\ee
which is the Lorentz force law on curved manifolds. 

A drawback of this hamiltonian formulation is, that the canonical momentum is not gauge invariant; under 
gauge transformations
\be
A'_{\mu} = A_{\mu} + \nb_{\mu} \Lb, \hs{2} p'_{\mu} = p_{\mu} + q \nb_{\mu} \Lb.
\label{5.3}
\ee
Therefore it is preferable to work with the covariant momentum \ct{gibbons-rietdijk-vholten1993,vholten2006}
\be
\pi_{\mu} = p_{\mu} - q A_{\mu} = m g_{\mu\nu} \dot{x}^{\nu},
\label{5.4}
\ee
in terms of which the hamiltonian takes the simple form 
\be
H = \frac{1}{2m}\, g^{\mu\nu} \pi_{\mu} \pi_{\nu}.
\label{5.5}
\ee
If the covariant derivative $\cD$ now is generalized to
\be
\cD_{\mu} G = \dd{G}{x^{\mu}} + \Gam_{\mu\nu}^{\;\;\;\lb}\, \pi_{\lb}\, \dd{G}{\pi_{\nu}}, 
\label{5.8}
\ee
then as before the metric postulate can be used to show that
\be 
\cD_{\mu} H = 0.
\label{5.10}
\ee
Moreover, the correct dynamics is reproduced as usual by
\be
\frac{dG}{d\tau} = \left\{ G, H \right\},
\label{5.7}
\ee 
when supplemented by the covariant brackets
\be
\left\{ G, K \right\} = \cD_{\mu} G\, \dd{K}{\pi_{\mu}} - \dd{G}{\pi_{\mu}}\, \cD_{\mu} K + q F_{\mu\nu}\, 
 \dd{G}{\pi_{\mu}} \dd{K}{\pi_{\nu}}.
\label{5.6}
\ee
In particular, this bracket reproduces the Ricci identity in the form
\be
\left\{ \pi_{\mu} , \pi_{\nu} \right\} = q F_{\mu\nu}.
\label{5.9}
\ee
In view of the identity (\ref{5.10}) and the anti-symmetry of the field strength tensor $F_{\mu\nu}$, the 
condition for a scalar quantity $G(x,\pi)$ to be constant of motion becomes \ct{vholten2006}
\be
\left\{ G, H \right\} = 0 \hs{1} \Rightarrow \hs{1} \pi^{\mu} \cD_{\mu} G = q \pi^{\mu} F_{\mu\nu}\, \dd{G}{\pi_{\nu}}.
\label{5.11}
\ee
In particular, if $G$ can be expressed as a power series
\be
G(x, \pi) = \sum_n \frac{1}{n!}\, G^{(n)\, \mu_1 ... \mu_n} (x)\, \pi_{\mu_1} ... \pi_{\mu_n}, 
\label{5.12}
\ee
then eq.\ (\ref{5.11}) takes the form of a first-order p.d.e.\ 
\be
\nb_{\lh \mu_1 \rd} G^{(n)}_{\ld \mu_2 ... \mu_{n+1} \rh} = q F_{\lh \mu_1 \rd}^{\;\;\nu}
 G^{(n+1)}_{\ld \mu_2 ... \mu_{n+1} \rh \nu}.
\label{5.13}
\ee
This is a generalization of equation (\ref{3.4}) for Killing tensors, forming a hierarchy of equation connecting 
tensors of different rank. Nevertheless, the existence of a Killing tensor of rank $n$ still provides a constant 
of motion: it allows the series expansion (\ref{5.12}) to be truncated, with all higher-rank components vanishing:
\be
\nb_{\lh \mu_1 \rd} G^{(n)}_{\ld \mu_2 ... \mu_{n+1} \rh} = 0 \hs{1} \Rightarrow \hs{1} G^{(n+k)}_{\mu_1 ... \mu_{n+k}} = 0,
\hs{1} \forall k \geq 0,
\label{5.14}
\ee
while all lower-rank components are obtained by solving eq.\ (\ref{5.13}) with the all-ready known higher-rank 
ones as inhomegenous source term on the right-hand side \ct{vholten2006}.
\vs{1}

\nit
Finally note, that it is straightforward to include a scalar potential $\Fg(x)$ as well:
\be
H = \frac{1}{2m}\, g^{\mu\nu} \pi_{\mu} \pi_{\nu} + \Fg, \hs{1} \Rightarrow \hs{1} \cD_{\mu} H = \dd{\Fg}{x^{\mu}}.
\label{5.15}
\ee
This modifies the generalization of the Killing equation to include a gradient term \ct{visinescu2009-2} 
\be
\pi^{\mu} \cD_{\mu} G = q \pi^{\mu} F_{\mu\nu} \dd{G}{\pi_{\nu}} + \dd{\Fg}{x^{\mu}} \dd{G}{\pi_{\mu}}.
\label{5.16}
\ee
{\em Example: a quantum-dot model} \\
A model of a quantum-dot, consisting of 2 electrons moving as a non-relativistic bound pair with Coulomb interaction
in a confining harmonic potential and a magnetic field, was studied in ref.\ \ct{cghhkz_2014}. Ignoring the center-of-mass 
motion, the pair can be described in axial co-ordinates $(\rg, z, \vf)$ as a single particle with hamiltonian
\be
H_{CM} = \frac{1}{2}\, g^{ij} \pi_i \pi_j + \Fg, 
\label{5.17}
\ee
where
\be
g_{ij} = \mbox{diag} (1, 1, \rg^2), \hs{2} 
\Fg = \frac{1}{2} \lh \og_0^2 \rg^2 + \og_z^2 z^2 \rh - \frac{\kg}{\sqrt{\rg^2 + z^2}}.
\label{5.18}
\ee
The definition of the dynamics is completed by the brackets 
\be
\left\{ G, K \right\} = \cD_i G \dd{K}{\pi_i} - \dd{G}{\pi_i} \cD_i K - 2 \og_L \rg \lh \dd{G}{\pi_{\rg}} \dd{K}{\pi_{\vf}} 
 - \dd{G}{\pi_{\vf}} \dd{K}{\pi_{\rg}} \rh,
\label{5.19}
\ee
where $\og_L = eB/2$ is the Larmor frequency. 

To guarantee that a Killing tensor can be completed to a full constant of motion, solving eq.\ (\ref{5.16}), it 
may be necessary to tune the parameters in the scalar potential. In the case of the quantum dot, there is
a rank-4 Killing tensor which can be completed to a full constant of motion provided the magnetic field is 
tuned to a value such that the Larmor frequency is a fixed combination of the harmonic frequencies:
\be
\og_L^2 + \og_0^2 = 4 \og_z^2.
\label{5.20}
\ee
Then the following expression is a solution of eq.\ (\ref{5.16}) for this system:
\be
\ba{l} 
G = \dsp{ \rg^2 \pi_z^4 - 2 \rg z \pi_{\rg} \pi_z^3 + z^2 \pi_{\rg}^2 \pi_z^2 + \frac{1}{\rg^2} \pi_{\vf}^4 + \pi_\rg^2 \pi_{\vf}^2 
   + \lh 2 + \frac{z^2}{\rg^2} \rh \pi_z^2 \pi_{\vf}^2 }\\ 
 \\
 \dsp{ +\, 2 \og_L \pi_{\vf} \lh \rg^2 \pi_{\rg}^2 + (2\rg^2 + z^2) \pi_z^2 \rh + \left[ 2 \og_z^2 z^3 \rg + \frac{2\kg z\rg}{\sqrt{\rg^2 + z^2}} \right] \pi_{\rg} \pi_z }\\
 \\
 \dsp{ + \left[ (2 \og_z^2 - \og_0^2) z^2 \rg^2 + 2 \og_L^2 \rg^4 - \frac{2\kg\rg^2}{\sqrt{\rg^2 + z^2}} \right] \pi_z^2 
  + \og_L^2\, \rg^4 \pi_{\rg}^2 }\\
 \\
 \dsp{ + \left[ 2 \og_z^2 z^2 + (\og_0^2 - 5 \og_L^2) \rg^2 - \frac{2\kg}{\sqrt{\rg^2 + z^2}} \right] \pi_{\vf}^2 }\\
 \\
 \dsp{ -\, 2 \og_L \pi_{\vf} \left[ (3 \og_L^2 - \og_0^2) \rg^4 - 2 \og_z^2 \rg^2 z^2 + \frac{2\kg\rg^2}{\sqrt{\rg^2 + z^2}} \right] 
 + \og_z^4 \rg^2 z^4 + 2 \og_z^2 \og_L^2 \rg^4 z^2  }\\
  \\
 \dsp{ - \og_L^2 ( 3 \og_L^2 - 4 \og_z^2) \rg^6 + \frac{2\kg}{\sqrt{\rg^2 + z^2}} \lh \og_z^2 \rg^2 z^2 - \og_L^2 \rg^4 \rh
  + \frac{\kg^2}{2}\, \frac{\rg^2 - z^2}{\rg^2 + z^2}. }
\ea
\label{5.21}
\ee

\section{Non-abelian gauge interactions}

The dynamics of a particle with non-abelian gauge interactions is described by Wong's generalization 
of the Lorentz force law \ct{wong1970}
\be
g_{\mu\nu} \lh \ddot{x}^{\nu} + \Gam_{\kg\lb}^{\;\;\;\nu} \dot{x}^{\kg} \dot{x}^{\lb} \rh = \frac{g}{m}\, t_a 
 F^a_{\mu\nu} \dot{x}^{\nu}, \hs{2} \dot{t}_a + g f_{ab}^{\;\;\;c} t_c A_{\mu}^{\;b} \dot{x}^{\mu}.
\label{6.1}
\ee
Here $g$ is the coupling constant, $f_{ab}^{\;\;\;c}$ are the structure constants of the Lie algebra of
gauge charges $t_a$, and the non-abelian field-strength tensor is
\be
F_{\mu\nu}^a = \nb_{\mu} A_{\nu}^a - \nb_{\nu} A_{\mu}^a + g f_{bc}^{\;\;\;a} A_{\mu}^b A_{\nu}^c.
\label{6.2}
\ee
The theory can be cast in hamiltonian form \ct{vholten2006} by introducing covariant momenta $\pi_{\mu}$ and a 
canonical hamiltonian
\be
H = \frac{1}{2m}\, g^{\mu\nu} \pi_{\mu} \pi_{\nu},
\label{6.3}
\ee
supplemented with brackets
\be
\ba{l} \dsp{ 
\left\{G, K \right\} = \cD_{\mu} G \dd{K}{\pi_{\mu}} - \dd{G}{\pi_{\mu}} \cD_{\mu} K 
 + g t_a F_{\mu\nu}^a \dd{G}{\pi_{\mu}} \dd{K}{\pi_{\nu}}
 + f_{ab}^{\;\;\;c} t_c\, \dd{G}{t_a} \dd{K}{t_b}, }\\
 \\
\dsp{ \cD_{\mu} G \equiv \dd{G}{x^{\mu}} + \Gam_{\mu\nu}^{\;\;\;\lb} \pi_{\lb} \dd{G}{\pi_{\nu}} 
 + g f_{ab}^{\;\;\;c}\, t_c\, A_{\mu}^a\, \dd{G}{t_b}. }
\ea
\label{6.4}
\ee
In particular these brackets guarantee the Ricci identity and the Lie-algebra of gauge charges:
\be
\left\{ \pi_{\mu}, \pi_{\nu} \right\} = g t_a\, F^a_{\mu\nu}, \hs{2}
\left\{ t_a, t_b \right\} = f_{ab}^{\;\;\;c} t_c.
\label{6.5}
\ee
It is now straightforward to derive the condition for the existence of constants of motion: 
\be
\left\{ G, H \right\} = 0 \hs{1} \Rightarrow \hs{1} \pi^{\mu} \cD_{\mu} G = g t_a F^a_{\mu\nu}\, \pi^{\mu} \dd{G}{\pi_{\nu}}.
\label{6.6}
\ee
In components this gives us a hierarchy of equations similar to (\ref{5.13}) for the abelian case:
\be
\ba{l}
\dsp{ \cD_{\mu} G^{(0)} = g t_a F_{\mu}^{a\, \nu} G^{(1)}_{\nu}, }\\
 \\
\dsp{ \cD_{\mu} G^{(1)}_{\nu} + \cD_{\nu} G^{(1)}_{\mu} = g t_a \lh F_{\mu}^{a\, \lb} G^{(2)}_{\lb\nu} 
 +  F_{\nu}^{a\, \lb} G^{(2)}_{\lb\mu} \rh, }\\
 \\
\dsp{ \cD_{\mu} G^{(2)}_{\nu\lb} + \cD_{\nu} G^{(2)}_{\lb\mu} + \cD_{\lb} G^{(2)}_{\mu\nu} = 
 g t_a \lh F_{\mu}^{a\, \kg} G^{(3)}_{\kg\lb\nu} +  F_{\nu}^{a\, \kg} G^{(2)}_{\kg\lb\mu} + F_{\lb}^{a\, \kg} G^{(3)}_{\kg\mu\nu} \rh, }\\
 \\
...
\ea 
\label{6.7}
\ee

\nit
{\em Example: 2-D $SU(2)$ Yang-Mills point charge} 
\vs{1}

\nit
As a simple example consider the dynamics of a non-abelian point charge in a static magnetic $SU(2)$-field 
in the euclidean plane. In the 2-dimensional plane such a magnetic field can be written as
\be
F^a_{ij} = \ve_{ij} B^a, 
\label{6.8}
\ee
where $a$ represents an adjoint vector component in 3-dimensional internal $SU(2)$-space. 
The free Yang-Mills equation then implies
\be
\nb_i B^a + g \ve^{abc} A^b_i  B^c = 0 \hs{1} \Rightarrow \hs{1} \nb_i B^{a\,2} = 0,
\label{6.9}
\ee
and the modulus of $B^a$ is constant. Now the direction of $B^a$ can be rotated point-wise in the plane by a 
local gauge transformation, and this freedom can be used to make $B^a$ constant. Such a constant $B^a$-field 
is derived from a vector potential
\be
A^a_i = - \frac{1}{2}\, \ve_{ij} x^j B^a.
\label{6.10}
\ee
Now the 2-dimensional euclidean plane is invariant under translations and rotations, guaranteeing conservation
of momentum and angular momentum. Our construction implies, that the angular momentum will obtain 
a field dependent addition:
\be
J = \ve_{ij} x_i \pi_j + \frac{g}{2}\, t_a B^a x_i^2.
\label{6.12}
\ee
In addition, there is also a pair of of rank-2 Killing tensors, generating a 
conserved Runge-Lenz vector:
\be
K_i =  x_i \pi_j^2 - \pi_i \pi_j x_j + g B^a t_a \lh \frac{1}{2}\, \ve_{ij} \pi_j x_k^2 + x_i x_j \ve_{jk} \pi_k \rh
 + \frac{1}{2}\, \lh g B^a t_a \rh^2 x_i\, x_j^2. 
\label{6.11}
\ee
\vs{2}

\nit
{\bf Acknowledgement} \\
The work described in this paper is supported by the Foundation for Fundamental Research of Matter (FOM),
as part of the research programme {\em Theoretical Particle Physics in the Era of the LHC}.

\end{document}